\documentclass[aip,twocolumn,jcp,groupedaddress]{revtex4}
\usepackage{amssymb}
\usepackage{amsmath}
\usepackage{dcolumn}
\usepackage{bm}
\usepackage{graphicx}
\usepackage{mathrsfs}
\usepackage{appendix}

\begin{document}

\title{Vibration-assisted coherent excitation energy transfer in a detuning system}
\author{Xin Wang, Hao Chen and Hong-rong Li$^{\dag}$}
\affiliation{Department of Applied Physics, Xi'an Jiaotong University, Xi'an 710049, China}

\begin{abstract}
The roles of the vibration motions played in the excitation energy transfer process are studied. It is found that a strong coherent transfer in the hybrid system emerges when the detuning between the donor and the acceptor equals the intrinsic frequency of the vibrational mode, and as a result the energy can be transferred into the acceptor much effectively. Three cases of the donor and the acceptor coupling with vibrational modes are investigated respectively. We find that the quantum interference between the two different transfer channels via the vibrational modes can affects the dynamics of the system significantly. \newline
$^{\dag}$Corresponding Email: hrli@mail.xjtu.edu.cn\newline
\end{abstract}

\maketitle

\ \ \ \ \ \ \ \ \ \ \ \ \ \ 

\section{Introduction}

Life on earth is sustained mainly by the energy of sunlight harvested through photosynthesis. It has been found that the energy absorbed by photosynthetic system can be transferred to reaction center (RC) with an efficiency higher than 95\% \cite{[1]}. Understanding the extreme high efficiency of the excitation energy transfer (EET) through coupled dipoles in the basic photosynthetic unit is one of big challenges in research on revealing the secret of this natural phenomenon \cite{[2],[3]}, and it also stimulates a motivation finding artificial bionic structures for energy storage and conservation, such as quantum dot nano-structures and polymer chain \cite{[4],[5],[6],[7]}.

No matter for an artificial or natural EET system, there generally exists three kind of interactions: First, within the bio-molecular system there will be couplings among different dipoles (sites) and these couplings are of dipole-dipole interaction form \cite{[8]}. Second, some dipoles will be excited after sequential dipole-dipole interaction and then dissipate the excitation energy into the RC \cite{[9]}, which is our expected process. Otherwise, the excited sites would dissipate to the thermal environment and induce the energy loss. Third, the dipoles will also interact with vibrational degrees caused by their intrinsic spatial structures (sites of linear, cyclic or other structures) and the motions induced by the thermal environment \cite{[10],[11]}. For example, in photosynthetic unit systems the pigments will be affected by the vibrations in the surrounding protein matrix and intra-chromophore vibrational modes \cite{[12],[13]}. To explore the role of vibrational modes played in the EET dynamics, many studies has been performed both in experimental and theoretical forms \cite{[14],[15],[16]}.

The basic unit of the EET systems includes a donor and an acceptor, each of which is usually treated as a two-level system (TLS). The two TLSs couple each other in the form of the Jaynes-Cummings (JC) model \cite{[8]}. In some cases the energy separations of the two TLSs are not the same (for example, B850 and B800 bacteriochlorophylls have absorption peaks at 850nm and 800nm respectively \cite{[17],[18]}), that is, energy transfer between the two TLSs is a detuning dynamic process. In some artificial systems, the transition frequencies for the donor and the acceptor may also not be the same as well \cite{[5]}. As is well known, energy can hardly be transferred between two TLSs with large detuning (compared with the JC coupling strength) without external assistance \cite{[19]}. Therefore, an important question is arisen, how does energy transfer happen with extremely high efficiency in the natural large detuning bio-systems?

In this work, we address to answer this question by theoretically exploring the energy transfer process in a detuning EET system assisted by the vibrational modes. To simplify the models we consider just energy transfer between one donor and one acceptor in the weak coupling regime \cite{[14]}, and study three cases: the basic model that only the donor couples with vibrational mode; the advanced model 1 that both the donor and the acceptor couple with the identical mode; the advanced model 2 that the two TLSs couples with different mode individually.

The paper is organized as follow: In section II, we show how to obtain the effective Hamiltonian of the basic model and discuss the coherent dynamics without any decoherent processes being considered. In section III, the numerical results of the basic model are presented. In section IV and V, investigations of the advanced model 1 and 2 are shown respectively. The conclusion is made in section VI.

\section{Theoretical analysis for the basic model}

In this section, we consider a basic model that includes a TLS1 (donor) and a TLS2 (acceptor), with energy separations $\epsilon _{1}$ and $\epsilon_{2} $ respectively. The interaction Hamiltonian for the two TLSs is ($\hbar =1$)

\begin{equation}
H_{D-A}=\epsilon _{1}\sigma _{1}^{+}\sigma _{1}^{-}+\epsilon _{2}\sigma_{2}^{+}\sigma _{2}^{-}+V_{12}\left( \sigma _{1}^{+}\sigma _{2}^{-}+\sigma_{2}^{+}\sigma _{1}^{-}\right) ,
\end{equation}
where $\sigma _{1}^{+}=\left\vert e_{1}\right\rangle \left\langle g_{1}\right\vert $ ($\sigma _{2}^{+}=\left\vert e_{2}\right\rangle \left\langle g_{2}\right\vert $) corresponds to transition operator on site 1(2) with energy detuning $\Delta =\epsilon _{1}-\epsilon _{2}$, and $V_{12}$ denotes the electronic coupling between the sites 1 and 2. Supposing that only the TLS1 (donor) is coupled to a vibrational mode with the free Hamiltonian $H_{M}=\upsilon b^{\dagger }b$, where $\upsilon $ is the angular eigenfrequency of the vibrational mode, and $b^{\dagger }$ ($b$) corresponds to the creation (annihilation) operator of the vibrational mode. The
coupling Hamiltonian between the TLS1 and the vibrational mode can be expressed as $H_{D-M}=g(b^{\dagger }+b)\sigma _{1}^{+}\sigma _{1}^{-}$, with $g$ corresponding to the coupling strength of the donor to the vibrational mode. Then the Hamiltonian describing the whole hybrid system has the form

\begin{equation}
H_{1}=H_{D-A}+H_{M}+H_{D-M}.
\end{equation}

Performing a unitary transformation $U_{0}(t)=\exp (iH_{0}t)$ to $H_{1}$ with $H_{0}=\epsilon _{2}(\sigma _{1}^{+}\sigma _{1}^{-}+\sigma_{2}^{+}\sigma _{2}^{-})$, and we obtain the interaction Hamiltonian as following

\begin{equation}
\begin{split}
H_{1I} = & \Delta \sigma _{1}^{+}\sigma _{1}^{-}+\upsilon b^{\dagger}b+V_{12}\left( \sigma _{1}^{+}\sigma _{2}^{-}+\sigma_{2}^{+}\sigma_{1}^{-}\right)\\ &+g(b^{\dagger }+b)\sigma _{1}^{+}\sigma _{1}^{-}.
\end{split}%
\end{equation}

As shown in appendix A, under the condition $\Delta =\upsilon $ and supposing that the coupling is weak, i.e., $\upsilon \gg \{g,V_{12}\}$, we can obtain the effective Hamiltonian under the Markov approximation: 
\begin{equation}
\begin{split}
H_{1,eff} = & (\frac{V_{12}^{2}}{\Delta }-\frac{g^{2}}{\Delta })\left\vert e_{1}\right\rangle \left\langle e_{1}\right\vert -\frac{V_{12}^{2}}{\Delta } \left\vert e_{2}\right\rangle \left\langle e_{2}\right\vert\\ &-\frac{V_{12}g}{\Delta }(b\sigma _{1}^{+}\sigma _{2}^{-}+b^{\dagger}\sigma _{1}^{-}\sigma _{2}^{+}).
\end{split}
\end{equation}

From Equation above, we find that the last term in the effective Hamiltonian (4) describing coherent transfer among the donor, the acceptor and the vibrational mode with rate $\frac{V_{12}g}{\Delta }$. We define two probabilities as $P_{i}(t)=$ Tr$[\left\vert e_{i}\right\rangle \langle e_{i}| \hat{\rho}(t)]$ ($i=1,2$) with $\hat{\rho}$ being the density matrix operator for the system. By supposing that the initial state of the system is $\left\vert e_{1},g_{2},n\right\rangle $, and no decoherent effects being considered, we plot the probabilities $P_{2}(t)$ to show mechanism of the energy transfer in our setup in Fig. 1(a), 1(b) and 1(c) that are determined
by $H_{1I}$ and $H_{1,eff}$ respectively, where $|n\rangle $ represents the Fock state of the vibrational mode with phonon number $n$. 
\begin{figure}[tbph]
\centering \includegraphics[width=9.0cm]{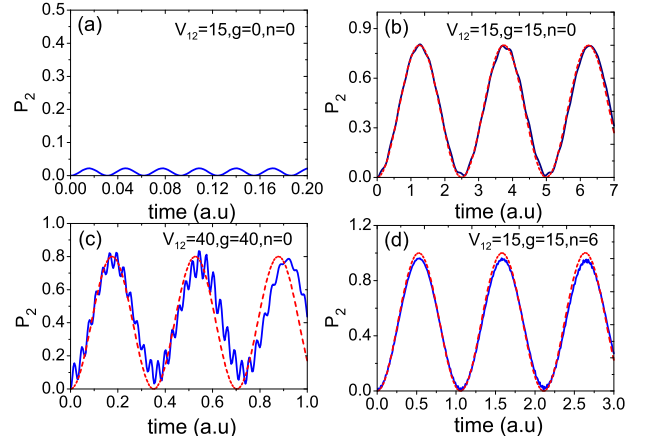}
\caption{The probabilities $P_{2}(t)$ verses time under different cases which are determined by Hamiltonian $H_{1I}$ (Blue solid line in (a), (b), (c), and (d)), $H_{1,eff}$ (red dashed line in (b), (c)) and $H_{1,eff}^{^{\prime}}$ (red dashed line in (d)), respectively. Here we set $\Delta =\protect\upsilon =200$.}
\label{fig1}
\end{figure}

In Fig. 1(a), we find that the excitation of the acceptor is in a very lower probability without the assistance of the vibrational mode because the detuning is much stronger than the coupling between the two TLSs. However, when the vibrational mode takes effects as shown in Fig. 1(b), the coherent transfer in this hybrid system emerges and the acceptor can be excited in a high probabilities. To understand this process more clearly, we give the energy level configuration that is shown in Fig. 2(a). We find that energy can be transferred from the donor to the acceptor (the red solid line and the brown solid line) with the phonon number increasing from $n$ to $n+1$, that is, the detuning energy $\Delta$ (the excess energy) between the donor and the acceptor leaks into the vibrational mode during the transition process. 
\begin{figure}[tbph]
\centering \includegraphics[width=8.0cm]{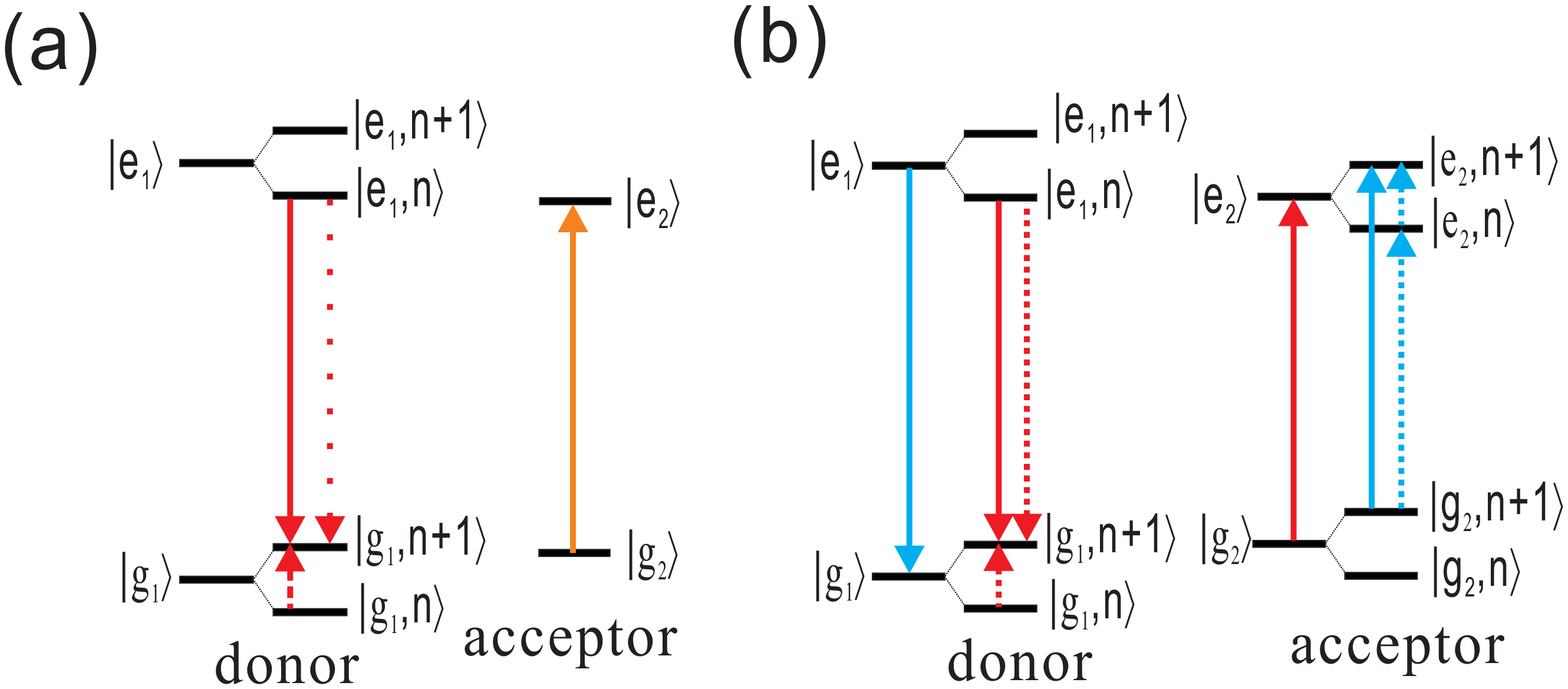}
\caption{Configurations of the energy levels: (a) the basic model; (b) the advanced model 1.}
\label{fig2}
\end{figure}

We show consistency of $H_{1,eff}$ and $H_{1I}$ under the weak coupling approximation by comparing Fig. 1(b) and Fig. 1(c). In Fig. 1(b), as the coupling strength $g$ is much weaker than the detuning $\Delta$, we find the approximation of obtaining the effective Hamiltonian $H_{1,eff}$ from $H_{1I}$ are valid. However, when the coupling strength becomes larger, difference of the figures can be observed as shown in Fig. 1(c), dynamics of probabilities $P_{2}(t)$ determined by $H_{1,eff}$ does not match so well with that determined by $H_{1I}$.

Moreover, as the effective spin-boson coupling strength is proportional to $\sqrt{n+1}$ \cite{[19]}, when the initial phonon number $n$ is very large, i.e. $\sqrt{n+1}\frac{V_{12}g}{\Delta }\gg \{\frac{V_{12}^{2}}{\Delta }- \frac{g^{2}}{\Delta },\frac{V_{12}^{2}}{\Delta }\}$, we can neglect the terms of energy shift in $H_{1,eff}$ and obtain a more simple effective Hamiltonian form 
\begin{equation}
H_{1,eff}^{^{\prime }}=-\frac{V_{12}g}{\Delta }(b\sigma _{1}^{+}\sigma_{2}^{-}+b^{\dagger }\sigma _{1}^{-}\sigma _{2}^{+}).
\end{equation}

The new form of the effective Hamiltonian $H_{1,eff}^{^{\prime }}$ includes only coherent transfer terms inside the system. As shown in Fig. 1(d), when $n=6$, the effects of energy shift terms which are contained in $H_{1,eff}$ are suppressed, and the evolution described by $H_{1,eff^{^{\prime }}}$ and $H_{1I}$ matches well, and $P_{2}(t)$ can reach nearly 1. We may point here that increasing phonon number at a certain range will enhance $P_{2}(t)$.

We may also explain mechanism of the coherent energy transfer of our system semi-classically. For the vibration-donor coupling term $H_{D-M}=g(b^{\dagger }+b)\sigma _{1}^{+}\sigma _{1}^{-}$, we can rewrite it as $H_{D-M}=F(t)x$, where $F(x)=\frac{gP_{1}(t)}{a_{0}},x=a_{0}(b^{\dagger }+b)$, and $a_{0}$ is the zero-point fluctuation of position amplitude for the vibrational mode \cite{[20]}. The new form of $H_{D-M}$ indicates that injection of an exciton of the donor creates a sudden force on the vibrational mode \cite{[13]}. Because the interaction of the donor and the acceptor is of large detuning, there must exist a Rabi oscillation with small amplitude for $P_{1}(t)$ at central frequency around $\Delta$ \cite{[19]}. We plot the distribution of the amplitude $A_{1}(\omega )$ ($\omega$ is the angular frequency at frequency domain) of the Fourier transform of $P_{1}(t)$ in Fig. (3), in which we know that there is indeed a peak around the eigenfrequency of the vibrational mode, i.e. $\omega =\upsilon =200$. It quite similar like that someone pushes a swing on its intrinsic frequency, although only an very weak force may be, the swing will oscillate with high amplitude after a relatively long time \cite{[21],[22]}. For this hybrid system, the excess energy $\epsilon _{1}-\epsilon _{2}=\Delta $ will leak into the vibrational mode resonantly. 
\begin{figure}[tbph]
\centering \includegraphics[width=8.0cm]{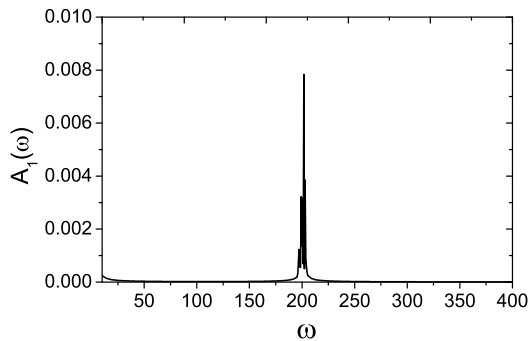}
\caption{The amplitude $A_{1}(\protect\omega )$ of Fourier transform for $P_{1}(t)$ at the frequency domain. The parameters adopted are same with those in Fig. 1(b).}
\label{fig3}
\end{figure}

\section{The numerical results of the basic model}

As demonstrated in the latest section, the vibrational mode can indeed assist coherent transfer between two TLSs without any decoherent effects being considered. However because that the environmental noise is inevitable, we may wonder whether the vibrational mode will also help to increase efficiency of energy transfer when the decay channels of the donor, acceptor and the vibrational mode take effects. In this section, we employ the numerical method to explore the role of the vibrational mode played in the hybrid dissipative system.

We assume that the system is polluted by three incoherent processes: two TLSs dissipate to the environment at rates $\Gamma _{1}$ and $\Gamma _{2}$, respectively; the acceptor can be sinked into the RC (sink) at a rate $\kappa$; moreover, the vibrational mode decays to thermal environment at a rate $\gamma $ with thermal phonon number $n_{th}\simeq kT/\hbar \upsilon$, where $k$ is the Boltzmann constant and $T$ is the temperature. All of the processes can be described by the Lindblad form terms $D[A,\Omega ]\hat{\rho}=\Omega (A\hat{\rho}A^{+}-\frac{1}{2}A^{+}A\hat{\rho}-\frac{1}{2}\hat{\rho}A^{+}A)$. Thus we can write the master equation of the whole system as follows 
\begin{equation}
\begin{split}
\frac{d\hat{\rho}(t)}{dt}= &-i\left[ H_{1,I},\hat{\rho}(t)\right] +D[\sigma_{1}^{-},\Gamma _{1}]\hat{\rho}(t)\\ & +D[\sigma _{2}^{-},\Gamma _{2}]\hat{\rho}(t)+D[\sigma _{2}^{-},\kappa ]\hat{\rho}(t)\\
& +n_{th}D[b^{+},\gamma ]\hat{\rho}(t)+(n_{th}+1)D[b,\gamma ]\hat{\rho}(t).
\end{split}
\end{equation}

We define a dissipation rate of the acceptor into the sink as $R(t)=\kappa \left\langle e_{2}\right\vert \hat{\rho}(t)|e_{2}\rangle $, then the efficiency at a fixed time $t$ can be calculated as $\eta(t)=\int_{0}^{t}R(t^{^{\prime }})dt^{^{\prime }}$ \cite{[23]}. We suppose that the TLSs are in $|e_{1},g_{2}\rangle$ initially, and our numerical simulation is based on the single excitation subspace thoroughly \cite{[24]}. To approach the actual environmental conditions, in all of the following numerical calculations, we assume that the vibrational mode has reached the thermal equilibrium states and the associated density matrix is $\mu(0)=\sum_{n}\frac{n_{th}^{n}}{(n_{th}+1)^{n+1}}\left\vert n\right\rangle \langle n|$. Thus the initial density matrix for the whole system can be expressed as $\hat{\rho}(0)=\left\vert e_{1},g_{2}\right\rangle \langle e_{1},g_{2}|\otimes \mu (0)$. We then perform the numerical simulations based on the master equation (6) and plot the time-dependent probabilities and the efficiency shown in Fig. 4.

Comparing Fig. 4(a) with Fig. 4(b), we find that evolution of the system have significant differences of the donor coupling or no coupling with the vibrational mode. When the donor is decoupled with the vibrational mode as shown in Fig. 4(b), it is found that $P_{2}(t) \simeq 0$, which means the acceptor is hardly excited and the efficiency is very low because of large detuning. In this case, most of the energy leaks out from the donor by dissipative processes and can not be transferred to the acceptor. However, when the vibrational mode takes part in which is shown in Fig. 4(a), the situation is totally different: the TLS2 can be excited with a high probability and the efficiency can reach above 80\%.

This can be understood as following. For the incoherent initial thermal state $\mu (0)=\sum_{n}\frac{n_{th}^{n}}{(n_{th}+1)^{n+1}}\left\vert n\right\rangle \langle n|$, we may first obtain $P_{2,n}(t)=$Tr$[\left\vert e_{2}\right\rangle \langle e_{2}|\hat{\rho}_{n}(t)]$ by solving the master equation (6) with the initial sate $\hat{\rho}_{n}(0)=\left\vert e_{1},g_{2},n\right\rangle \langle e_{1},g_{2},n|$ ($n=1,2,3...$) individually for different $n$, because $\mu (0)$ is only an incoherent mixed state with different Fock states, there is no coherence between $\rho_{n}(t)$, and then we can obtain weighted average $P_{2}(t)=\sum_{n}\frac{n_{th}^{n}}{(n_{th}+1)^{n+1}}P_{2,n}(t)$. For each $n$, when the vibrational mode is coupled with the EET system, there will be a coherent transfer
between state $\left\vert e_{1},g_{2},n\right\rangle $ and $\left\vert g_{1},e_{2},n+1\right\rangle $. It means that the energy can be transferred to the acceptor and as a result the phonon number of the vibrational mode increasing from $n$ to $n+1$ (as shown in Fig. 2(a)). Directly energy transfer from TLS1 to TLS2 without helpfulness of the vibrational mode is not an energy conservation process. However the transfer between state $\left\vert e_{1},g_{2},n\right\rangle $ and $\left\vert g_{1},e_{2},n+1\right\rangle $ with the assistance of vibrational mode satisfies energy conservation ($\epsilon _{1}=\epsilon _{2}+\upsilon )$, thus it can happen with a high probability.

\begin{figure}[tbph]
\centering \includegraphics[width=8.0cm]{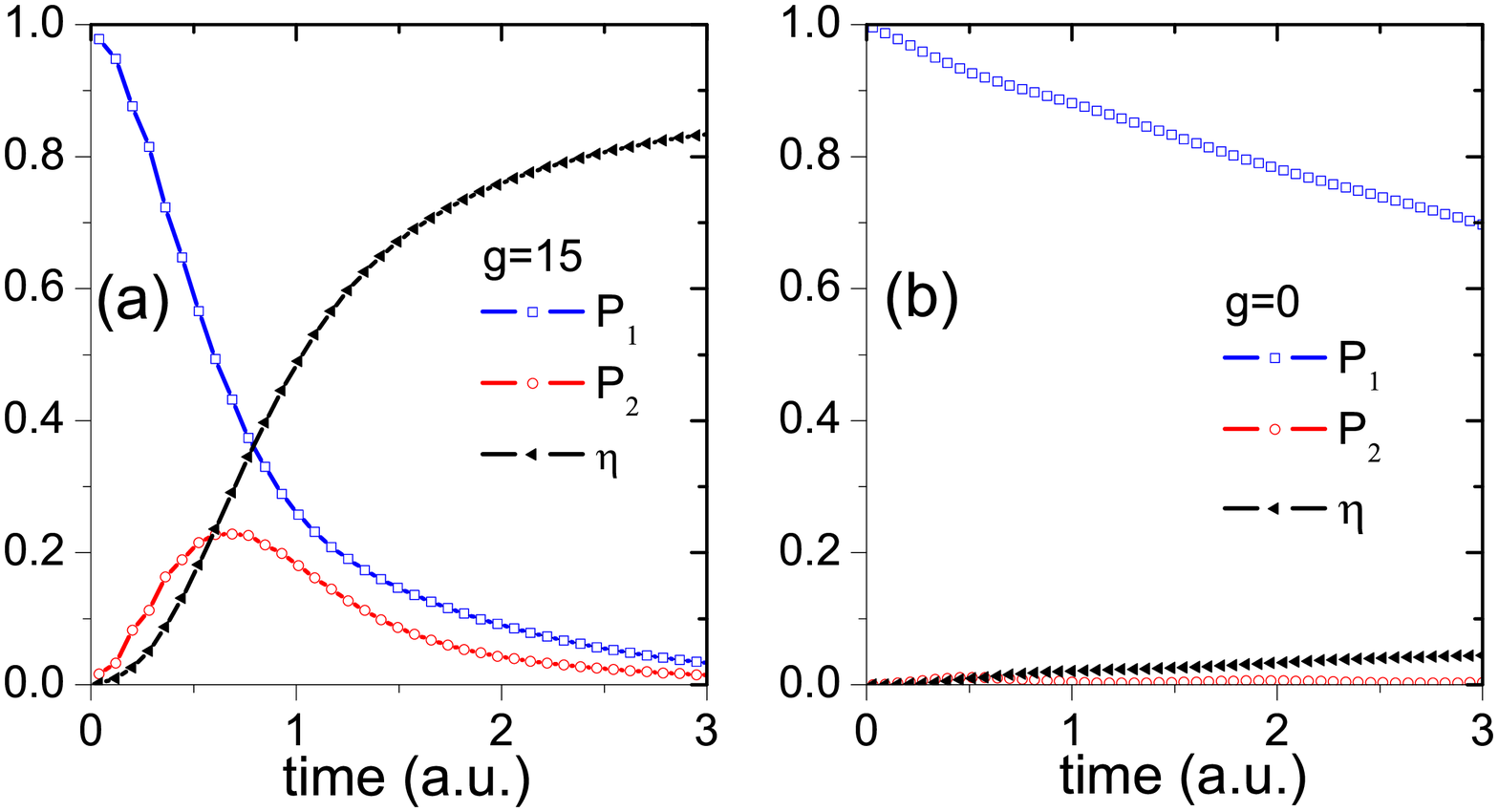}
\caption{Numerical results of the probabilities $P_{1}(t),$ $P_{2}(t)$ and the efficiency $\protect\eta (t)$ changes with time. We choose $\Delta =\protect\upsilon =200,V_{12}=15$, $\Gamma _{1}=\Gamma _{2}=0.1$, $\protect\kappa =3$, $\protect\gamma =0.4$, $n_{th}=1$.}
\end{figure}

We demonstrate the numerical results in Fig. 5 with adapting different initial thermal states. From Fig. 5(a) we find that, compared with the case when $n_{th}=1$, both the transfer efficiency $\eta (t)$ and the probability $P_{2}(t)$ is higher when $n_{th}=6$, and the system reaches its steady state more faster. When the average phonon number $n_{th}$ of the initial state becomes larger, the effective coupling strength become stronger, and as a result the energy can be transferred to the acceptor with a higher efficiency in a shorter period of time. Fig. 5(b) displays that the efficiency $\eta (t)$ increases with $n_{th}$ under different coupling strength $g$ when the system reaches its steady state. We find that the impact of increasing $n_{th}$ is apparent when the coupling is relatively weak ($g=5$), however the effect is not so well when the coupling is much stronger (for example $g=15$).

\begin{figure}[tbph]
\centering \includegraphics[width=8.0cm]{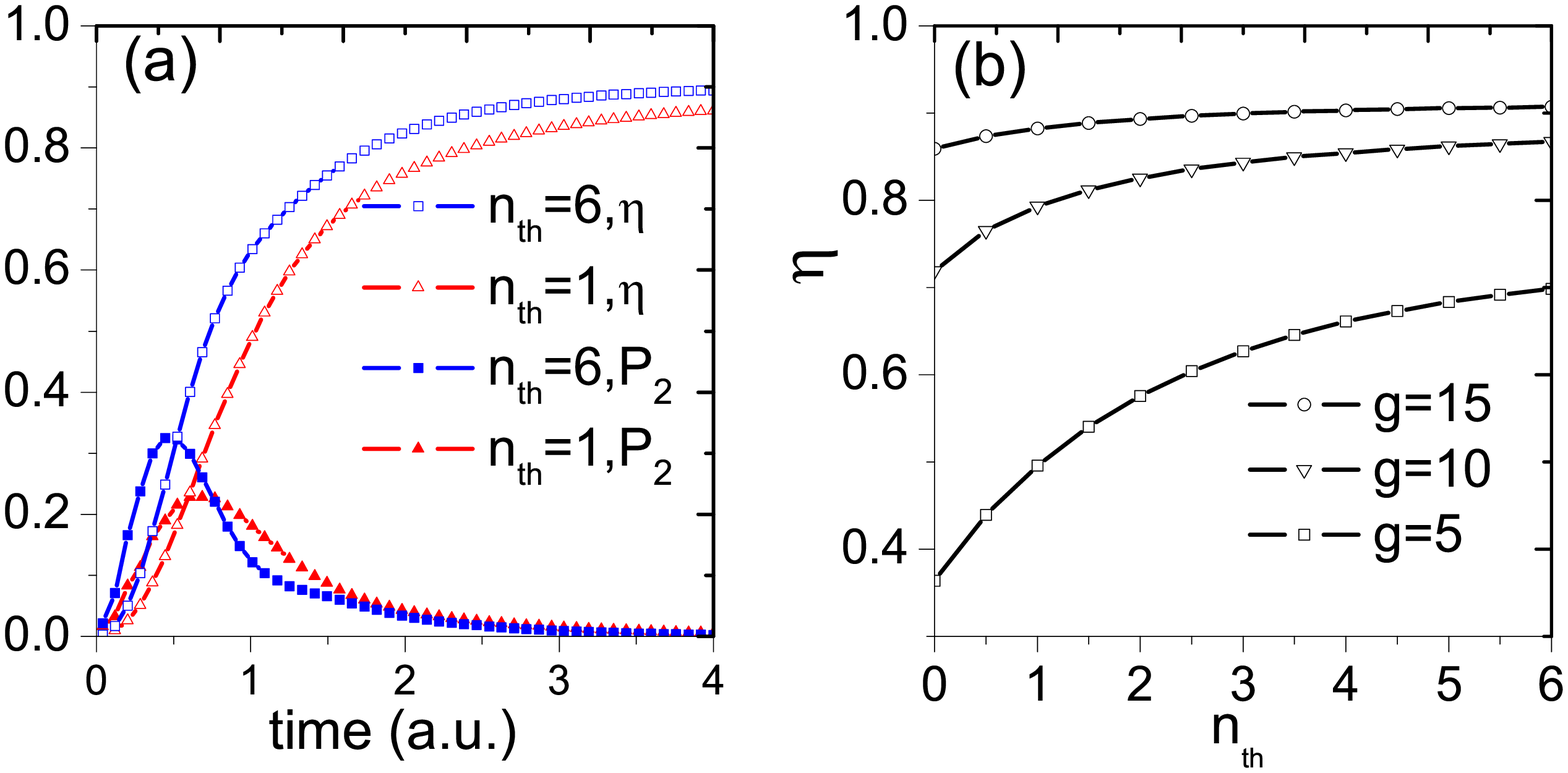}
\caption{(a) The evolution of $\protect\eta (t)$ and $P_{2}(t)$ with $n_{th}=1$ and $n_{th}=6$, and the parameters is the same as that in Fig. 4(a). (b) $\protect\eta $ as a function of $n_{th}$ for different coupling strength $g=5$, $g=10$ and $g=15$ when $t=10$ (with ensuring the system has reached its steady state).}
\end{figure}

\section{The advanced model 1: Both of the two TLSs are coupling with the identical vibrational mode}

We have shown that vibration mode can indeed assist the energy transfer between the two detuning TLSs with only the donor coupling the vibrational mode. Here we want to explore another case, in which the acceptor and the donor both couple with the identical vibrational mode. The Hamiltonian describes the system reads

\begin{equation}
H_{2}=H_{D-A}+H_{M}+H_{D-M-A},
\end{equation}
where $H_{D-M-A}=(g_{1}b^{\dagger }+g_{1}^{\ast }b)\sigma _{1}^{+}\sigma_{1}^{-}+(g_{2}b^{\dagger }+g_{2}^{\ast }b)\sigma _{2}^{+}\sigma _{2}^{-}$, and $g_{1}$ and $g_{2}$ are the coupling strength of the vibrational mode for the donor and the acceptor, respectively. We suppose $g_{1}$ is real and $g_{2}=|g_{2}|e^{i\theta }$, and $\theta $ is the relative phase of $g_{1}$ and $g_{2}$. With supposing $\Delta =\upsilon \gg \{g_{1},g_{2},V_{12}\}$, we obtain an effective Hamiltonian for the system

\begin{equation}
\begin{split}
H_{2,eff}= & (\frac{V_{12}^{2}}{\Delta }-\frac{g_{1}^{2}}{\Delta })\left\vert e_{1}\right\rangle \left\langle e_{1}\right\vert -(\frac{V_{12}^{2}}{\Delta }+\frac{|g_{2}|^{2}}{\Delta })\left\vert e_{2}\right\rangle \left\langle
e_{2}\right\vert \\ &-\frac{2g_{1}|g_{2}|}{\Delta }\cos \theta \left\vert e_{1}\right\rangle \left\langle e_{1}\right\vert \otimes \left\vert e_{2}\right\rangle
\left\langle e_{2}\right\vert \\ &+\frac{V_{12}(g_{2}^{\ast }-g_{1})}{\Delta }b\sigma _{1}^{+}\sigma_{2}^{-}+\frac{V_{12}(g_{2}-g_{1})}{\Delta }b^{\dagger }\sigma
_{1}^{-}\sigma _{2}^{+}.
\end{split}
\end{equation}

The first three terms in $H_{2,eff}$ are the energy shifts caused by the couplings among the three sub-systems. The laster two terms in $H_{2,eff}$ represent coherent transfer with rate $\frac{V_{12}(g_{2}-g_{1})}{\Delta }$, which shows that the difference of the two coupling strengths has a significant impact on the coherent transfer process. As shown in Fig. 2(b), there are two transition channels: The excess energy $\Delta $ can leak into the vibrational mode via coupling with the donor or with the acceptor at rate $\frac{-V_{12}g_{1}}{\Delta }$ (red line) or $\frac{V_{12}g_{2}}{\Delta }$ (blue line), respectively.

By considering the decoherent process similar as the basic model $P_{2}(t)$ and $\eta (t)$ are plotted in Fig. 6 based on the numerical calculation. 
\begin{figure}[tbph]
\centering \includegraphics[width=9.0cm]{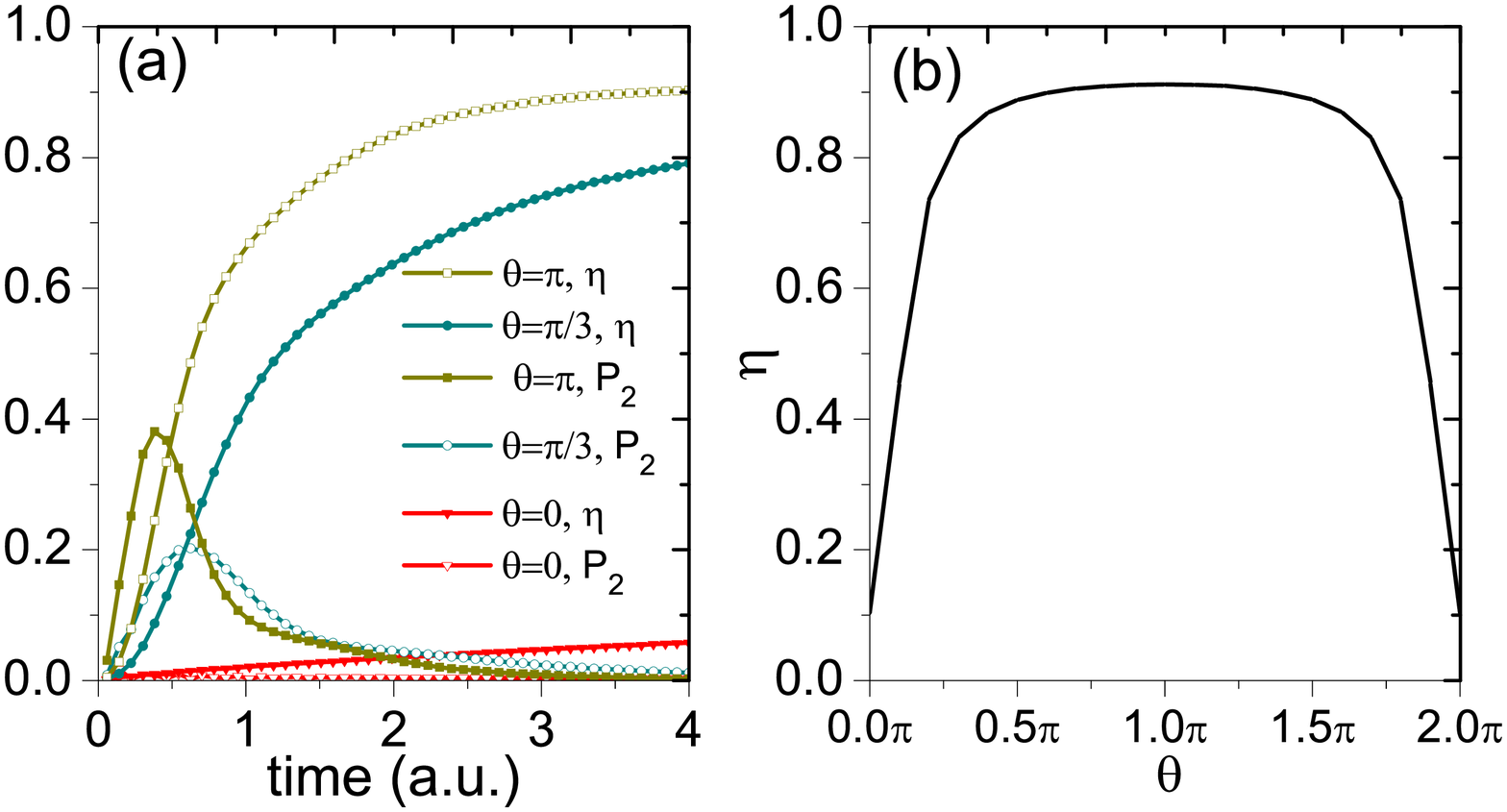}
\caption{Numerical results for the advanced model 1 with: (a) $\protect\eta (t)$ and $P_{2}(t)$ versus time for $\protect\theta =0,\protect \theta =\protect\pi /3$ and $\protect\theta =\protect\pi $; (b) $\protect \eta $ as a function of $\protect\theta $ when $t=10$. The parameters is the same as that in Fig. 4(a) except $g_{1}=|g_{2}|=15.$}
\label{fig6}
\end{figure}
In Fig. 6(a), we consider three cases with $\theta =0$, $\theta =\pi /3$ and $\theta =\pi $, respectively. We find that the relative coupling phase of $g_{1}$ and $g_{2}$ affects the dynamic of the system significantly. Comparing with the case of $\theta =\pi /3$, for the case of $\theta =\pi$, the efficiency $\eta (t)$ arises faster and the peak of $P_{2}(t)$ is of much higher and is coming more earlier. Correspondingly, the steady $\eta$ of the case with $\theta =\pi $ is higher than that of the case with $\theta =\pi /3$. Total differently, the situation of $\theta =0$ is quite similar with the case of no assistance of the vibrational mode.

We plot the dependency of steady $\eta $ on $\theta $ in Fig. 6(b). We find that the steady $\eta $ varies greatly with the relative coupling phase, which can be clearly understood. There are two energy transition channels as shown in Fig. 2(b), and these two channels interfere each other through coupling the identical vibrational mode, and the inter-coupling rate (strength) is $\frac{V_{12}(g_{2}-g_{1})}{\Delta }$. The coupling phase $\theta $ will take effects in determining whether the quantum interference between two channels is destructive or constructive. As $\frac{V_{12}(g_{2}e^{i\theta }-g_{1})}{\Delta }=0$ when $\theta =0$ and $|g_{2}|=g_{1}$, energy can hardly be transferred from the donor to the acceptor under this conditions. However when $\theta =\pi $ the quantum interference of two channel becomes constructive and the coherent transfer rate is $\frac{2V_{12}g_{1}}{\Delta }$, which makes energy can be transfer into the acceptor with a high probability.

We give also the semiclassical explanation from $H_{D-M-A}$ of the advanced model 1 when $\theta =0$ and $\theta =\pi $. The Hmiltonian can be rewritten as $H_{D-M-A}=(F_{1}(t)+F_{2}(t))x$, where $F_{i}(t)=\frac{g_{i}P_{i}(t)}{a_{0}}$ ($i=1,2)$. Different from that in the basic model, these two forces $F_{1}(t)$ and $F_{2}(t)$ are induced by the donor and acceptor, respectively. In Fig. 7 we plot the amplitude image $A_{i}(\omega )$ and phase image $\Theta _{i}(\omega )$ of Fourier transform of $P_{i}(t)$ ($i=1,2)$ and calculate the phase difference $\Theta (\omega )=\Theta_{1}(\omega )-\Theta _{2}(\omega )$ in the frequency domain, where the parameters adopted are same with those in Fig. 1(b) except $g_{1}=|g_{2}|=15$.
\begin{figure}[tbph]
\centering \includegraphics[width=9.0cm]{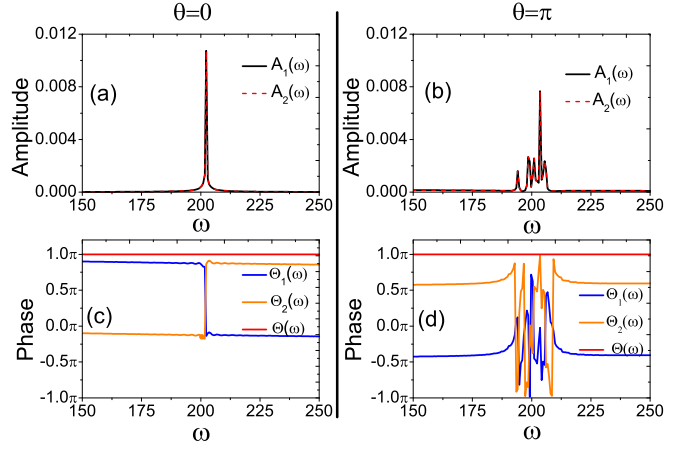}
\caption{(a) and (b) show amplitude images $A_{1}(\protect\omega )$ (black line) and $A_{2}(\protect\omega )$ (red dashed line) when $\protect\theta =0$ and $\protect\theta =\protect\pi$, respectively. (c) and (d) show phase image $\Theta _{1}(\protect\omega )$ (blue line), $\Theta _{2}(\protect \omega )$ (brown line) and $\Theta (\protect\omega )$ (red line) when $\protect\theta =0$ and $\protect\theta =\protect\pi$, respectively.}
\label{fig7}
\end{figure}

From Fig. 7(a) and Fig. 7(b), we find that the peaks of $A_{1}(\omega )$ and $A_{2}(\omega )$ locate at $\omega =\upsilon =200$ around no matter for $\theta =0$ or $\theta =\pi $. However, the phase differences $\Theta (\omega)$ always equal $\pi $ (shown in Fig. 7(c) and Fig. 7(d)), that means that $P_{1}(t)$ and $P_{2}(t)$ are of opposite sign all the time. When $\theta =0$, the forces induced by the donor and acceptor are opposite and as a result the energy can hardly leak into the vibrational mode. For the case of $\theta =\pi $, the total relative phase of $F_{1}(t)$ and $F_{2}(t)$ equals $0$, it indicates that the direction of the forces induced by the donor and acceptor are the same and thus the excess energy can effectively leak into the vibrational mode.

\section{The advanced model 2: Two TLSs are coupling with different vibrational modes}

We now discuss a more complex case: the donor and the acceptor are coupled with different vibrational modes, respectively. The Hamiltonian can be expressed as
 
\begin{equation}
\begin{split}
H_{3}& =H_{D-A}+H^{\prime}_{M}+H_{D-M_{12}-A},
\end{split}
\end{equation}
where $H^{\prime}_{M}=\upsilon _{1}b_{1}^{\dagger}b_{1}+\upsilon_{2}b_{2}^{\dagger }b_{2}$, and $H_{D-M_{12}-A}=g_{1}^{\ast
}b_{1})\sigma_{1}^{+}\sigma _{1}^{-}+(g_{2}b_{2}^{\dagger }+g_{2}^{\ast}b_{2})\sigma _{2}^{+}\sigma _{2}^{-}$, and $\upsilon_{1}$ and $\upsilon_{2}$ are the angular eigenfrequencies for mode 1 and mode 2 respectively, and $b_{i}^{\dagger }$ ($b_{i}$) corresponds to the creation (annihilation) operator of the vibrational mode $i$, and $g_i$ represents the coupling strength of the donor ($i=1$) or the acceptor ($i=2$) to the associated vibrational mode. We suppose $g_{1}$ is real and $g_{2}=|g_{2}|e^{i\theta }$, where $\theta$ is the relative coupling phase. Under $\Delta=\upsilon_{1}=\upsilon _{2}\gg \{g_{1},g_{2},V_{12}\}$, we obtain the effective Hamiltonian for $H_{3}$ as below 
\begin{equation}
\begin{split}
H_{3,eff} = &(\frac{V_{12}^{2}}{\Delta }-\frac{g_{1}^{2}}{\Delta })\left\vert e_{1}\right\rangle \left\langle e_{1}\right\vert -(\frac{V_{12}^{2}}{\Delta}+\frac{|g_{2}|^{2}}{\Delta })\left\vert e_{2}\right\rangle \left\langle e_{2}\right\vert \\ & -\frac{V_{12}g_{1}}{\Delta }(b_{1}\sigma _{1}^{+}\sigma_{2}^{-}+b_{1}^{\dagger }\sigma _{1}^{-}\sigma _{2}^{+}) \\ & +\frac{V_{12}}{\Delta }(g_{2}^{\ast }b_{2}\sigma _{1}^{+}\sigma_{2}^{-}+gb_{2}^{\dagger}\sigma _{1}^{-}\sigma _{2}^{+}).
\end{split}
\end{equation}

Apparently there are also two transition channels for energy transfer from the donor to the acceptor: energy can be transferred from the donor to the acceptor with creation of a phonon in mode 1 (being coupled with the first TLS) or in mode 2 (being coupled to the second TLS). That is, the excess energy $\Delta $ leaks into the vibrational mode 1 or mode 2, which is similar as advanced model 1. However, these two channels are independent because the two modes are independent. Thus there is no coherent cancellation and the relative coupling phase $\theta $ does not affect the evolution of the whole system, which can be proved by the numerical results in Fig. 8 with choosing $\theta =0$ and $\theta =\pi $. Where we suppose the initial state is $\hat{\rho}(0)=\left\vert e_{1},g_{2}\right\rangle \langle e_{1},g_{2}|\otimes \mu _{1}(0)\otimes \mu _{2}(0)$, where $\mu _{1}(0)$ and $\mu _{2}(0)$ are the the thermal states for the vibrational mode 1 and vibrational mode 2 with $n_{th}=1$. 
\begin{figure}[tbph]
\centering \includegraphics[width=8.0cm]{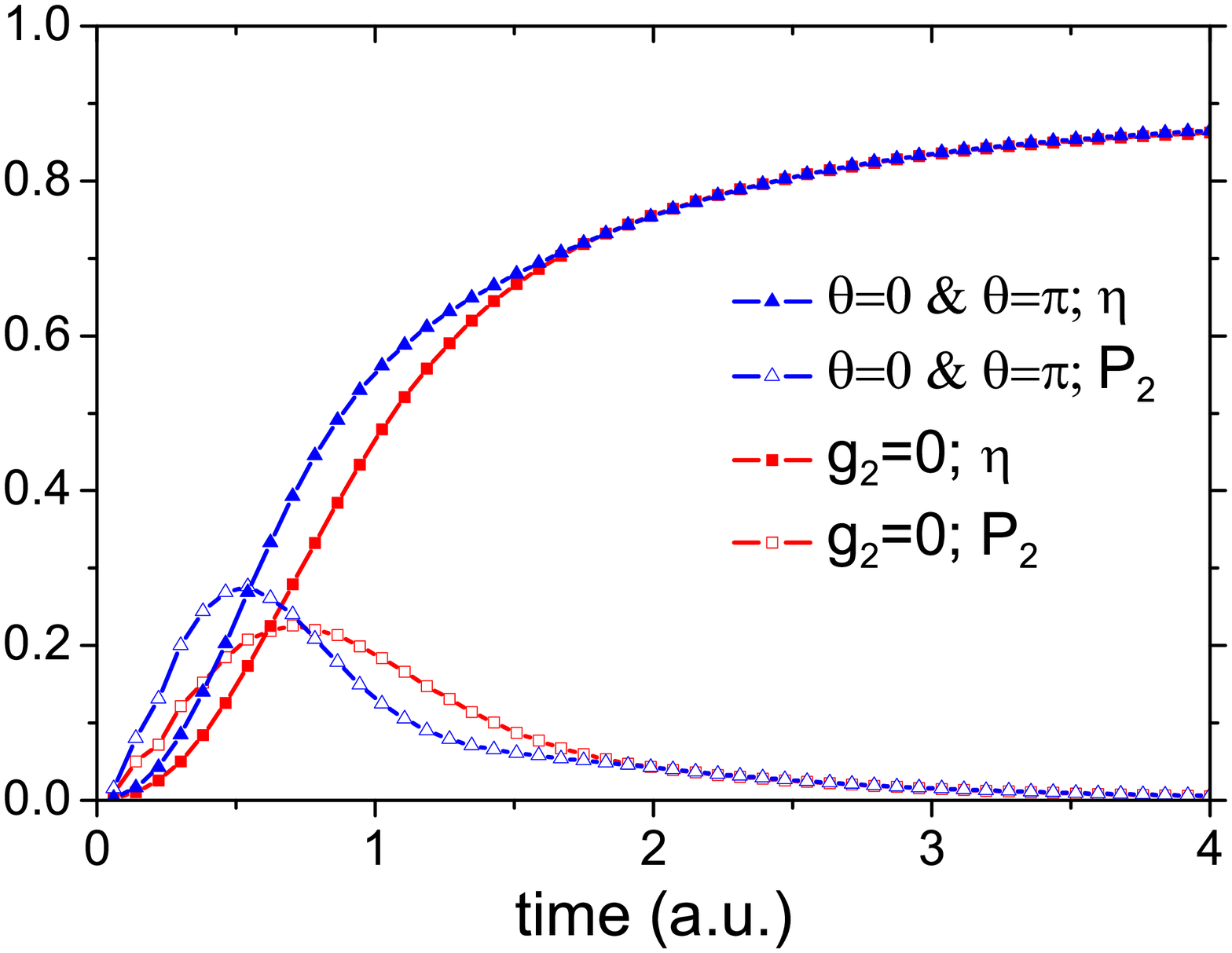}
\caption{The evolutions of the advanced model 2. Parameters adopted here is the same as those in Fig. 4(a) except $g_{1}=|g_{2}|=15$ of the blue line, while $g_{1}=15$, and $g_{2}=0$ of the red line.}
\label{fig8}
\end{figure}
The evolution of $P_{2}(t)$ and $\eta (t)$ under $\theta =0$ and $\theta=\pi $ coincide each other (blue lines marked with triangles) as shown in Fig. 8. Moreover, because both two channels take effects, the efficiency $\eta (t)$ gets larger more quickly and the steady value is slightly higher than the case that $g_{2}=0$ (red lines marked with squares), which corresponds to the basic model discussed in section III.

\section{conclusion}

In summary, we have studied the energy transfer process between two detuning TLSs assisted by the vibrational modes. By analyzing the effective Hamiltonian of the system, we found that there exists second order coherent transfer channels in these hybrid systems. The excess energy (detuning between two TLSs) can leak into the vibrational modes via these channels, and thus both coherence and the EET efficiency can be improved significantly. By supposing the vibrational modes are initially in incoherent thermal states, we find the EET efficiency is higher in a relatively high temperature environment. We also find that the quantum interference between two channels plays an important role in the evolution of the system and affects the final EET efficiency greatly. Our results may open up experimental possibilities to investigate and explore detuning coherent transfer phenomena in artificial and natural excitation energy transfer systems.

\section*{Acknowledgments}

This work is supported by the Natural Science Foundation of China under
Grant No. 11174233.

\setcounter{equation}{0} \renewcommand{\theequation}{A\arabic{equation}}

\setcounter{figure}{0} \renewcommand{\thefigure}{A\arabic{figure}}

\begin{appendix}

\section{Steps of getting the effective Hamiltonian $H_{1, eff}$}

We start
form $H_{1I}$. Performing another unitary transformation $U_{A1}(t)=\exp (iH_{A1}t)$ to $H_{1I}$ with $H_{A1}=\Delta \sigma
_{1}^{+}\sigma_{1}^{-}+\upsilon b^{\dagger }b$, then we get

\begin{equation}
\begin{split}
H_{AI}(t)&=V_{12}\left( \sigma _{1}^{+}\sigma _{2}^{-}e^{i\Delta t}+\sigma_{2}^{+}\sigma _{1}^{-}e^{-i\Delta t}\right) \\
&+g(b^{\dagger}e^{i\upsilon t}+be^{-i\upsilon t})\sigma _{1}^{+}\sigma _{1}^{-}.
\end{split}
\end{equation}

With supposing $\Delta =\upsilon $, $H_{AI}(t)$ can be rewritten as

\begin{equation}
H_{AI}(t)=Be^{i\Delta t}+B^{\dagger }e^{-i\Delta t},
\end{equation}

where $B=(V_{12}\sigma _{1}^{+}\sigma_{2}^{-}+gb^{\dagger}\sigma_{1}^{+}\sigma _{1}^{-})$.

Assuming that the detuning $\Delta$ is much larger than the coupling strength $\{g,V_{12}\}$, and focusing on the evolution over a period much longer than the period of any of the oscillations of the system, we can obtain the effective Hamiltonian expressed as \cite{[25]}
\begin{equation}
\begin{split}
H_{A1,eff} &=\frac{1}{\Delta }(BB^{\dagger }-B^{\dagger }B) \\
&=\frac{V_{12}^{2}}{\Delta }(\left\vert e_{1}g_{2}\right\rangle \left\langle e_{1}g_{2}\right\vert -\left\vert e_{2}g_{1}\right\rangle \left\langle e_{2}g_{1}\right\vert )\\
&-\frac{g^{2}}{\Delta }\left\vert e_{1}\right\rangle\left\langle e_{1}\right\vert-\frac{V_{12}g}{\Delta }(b\sigma_{1}^{+}\sigma _{2}^{-}+\sigma _{1}^{-}\sigma _{2}^{+}).
\end{split}
\end{equation}
The first term in $H_{A1,eff}$ represents the energy shift caused by the coupling between the donor and the acceptor, and the second term represents the energy shift caused by the interaction between the vibrational mode and the donor. Using the relation $\left\vert e_{i}\right\rangle \left\langle e_{i}\right\vert+\left\vert g_{i}\right\rangle \left\langle g_{i}\right\vert =I(i=1,2),$ we then rewrite $H_{A1,eff}$ as

\begin{equation}
\begin{split}
H_{A1,eff}&=(\frac{V_{12}^{2}}{\Delta }-\frac{g^{2}}{\Delta })\left\vert e_{1}\right\rangle \left\langle e_{1}\right\vert -\frac{V_{12}^{2}}{\Delta }\left\vert e_{2}\right\rangle \left\langle e_{2}\right\vert \\
&-\frac{V_{12}g}{\Delta }(b\sigma _{1}^{+}\sigma _{2}^{-}+b^{\dagger }\sigma _{1}^{-}\sigma_{2}^{+}).
\end{split}
\end{equation}

\end{appendix}

\end{document}